\newcommand{\CLASSINPUTtoptextmargin}{0.8in}

\documentclass[conference]{IEEEtran}

\usepackage[cmex10]{amsmath}
\usepackage{amsmath}
\usepackage{amssymb}
\usepackage{graphicx}
\usepackage{algorithm,algorithmic}
\usepackage[algo2e]{algorithm2e}

\begin{document}

\title{Extremely Large Aperture Massive MIMO:\\ Low Complexity Receiver Architectures}
\author{
\IEEEauthorblockN{Abolfazl Amiri,\\ Marko Angjelichinoski, Elisabeth de Carvalho  }\\
\IEEEauthorblockA{ Department of Electronic Systems\\ Aalborg University, Denmark\\
Email: \{aba,maa,edc\}@es.aau.dk}
\and
\IEEEauthorblockN{Robert W.\ Heath Jr.\  } \\
\IEEEauthorblockA{ Department of Electrical and Computer Engineering\\
The University of Texas at Austin, TX, USA\\
Email:  rheath@utexas.edu}}

\maketitle
%\clearpage\maketitle
\thispagestyle{empty}
\date{}

\newpage
\begin{abstract}
This paper focuses on new communication paradigms arising in massive multiple-input-multiple-output systems where the antenna array at the base station is of {extremely large dimension} (xMaMIMO).
Due to the extreme dimension of the array, xMaMIMO is characterized by spatial non-stationary field properties along the array; this calls for a multi-antenna transceiver design that is adapted to the array dimension but also its non-stationary properties. 
% In particular, due to the large signal attenuation along the array, one given user effectively communicates with only one portion of the array, defined as a visibility region. 
We address implementation aspects of xMaMIMO, with computational efficiency as our primary objective.
To reduce the computational burden of centralized schemes, we distribute the processing into smaller, disjoint sub-arrays.
Then, we consider several low-complexity data detection algorithms as candidates for uplink communication in crowded xMaMIMO systems.
Drawing inspiration from coded random access, one of the main contributions of the paper is the design of low complexity scheme that exploits the non-stationary nature of xMaMIMO systems and where the data processing is decentralized.
We evaluate the bit-error-rate performance of the transceivers in crowded xMaMIMO scenarios. The results confirm their practical potential. 
\end{abstract}

\begin{IEEEkeywords}
Very large arrays, Massive MIMO, coded random access, non-stationary, 5G
\end{IEEEkeywords}

\section{Introduction}

%Massive multiple-input-multiple-output (MIMO) is an assuring solution for next generation of mobile communications. The main idea is to use a massive number of antennas, i.e. hundreds, in a base station to provide spectral as well as energy efficiency \cite{petmag},\cite{larsson_mimo}. Having many more antennas in the base station than the user devices provides quasi-orthogonal channels for each user. Favorable propagation \cite{larsson_energy} helps in reducing randomness in the channels and achieving near-optimal gains.

%Concept of very large antennas (VLAs) is a new solution for nowadays increased wireless communication demands \cite{Husha},\cite{russek}. They  can be seen as an extension for massive MIMO systems with scaled up antenna elements. This large array can support more users in addition to higher quality of services. On the other hand, VLAs require enormous computational load which is one the most important challenges in their practical implementation. 

Massive multiple-input-multiple-output (MIMO) is a key technology in cellular communication systems for increasing area spectral efficiency \cite{petmag}, \cite{larsson_mimo}. The highest gains with massive MIMO are achieved when the antenna array dimension is very large~\cite{Lund, measurement_paper}. 
This has motivated the introduction of new types of deployment where arrays with extremely large dimension are deployed as part of a large infrastructure, for example along the walls of buildings in a mega-city, in airports, large shopping malls or along the structure of a stadium~\cite{measurement_paper}. Similarly, large intelligent surfaces have emerged involving large electromagnetic surfaces \cite{beyond_mimo}.  Such a massive MIMO system with antenna arrays of extremely large dimension is denoted as xMaMIMO.
%as the array dimension characterizes the number of spatial degrees of freedom available for communication. This motivated the introduction of new types of deployment where arrays with very large dimension are deployed as part of a large infrastructure, for example along the walls of buildings in a mega-city, in airports, large shopping malls or along the structure of a stadium \cite{metis}. Such extremely large arrays and their properties were first investigated through channel measurements~\cite{measurement_paper, AAU2, Lund}. Recently, large intelligent surfaces have emerged involving large electromagnetic surfaces and similar concepts for communications \cite{beyond_mimo, Husha}. 

With increased antenna array dimensions, spatial non-wide sense stationary properties appear across the array due to electromagnetic propagation attributes as well as the distance between the users and the array that becomes smaller than the Rayleigh distance (see Fig. 1). In such xMaMIMO systems, different channel models and receiver algorithms are needed that account for this non-stationarity. 

In this paper, we consider non-stationary properties through the concept of visibility region. A visibility region is associated to one given user and is defined as the portion of the array that one given user sees, i.e. that is able to receive signals from the user.
This behaviour of the channel introduces an inherent \emph{sparsity} to the system model, meaning that the transmitted signal of one user only exists on a small part of the antenna array. Thus, in contrast to ordinary massive MIMO models, user detection can be done by only processing the visibility region of each user. Using this important property of the system, we cut the computation costs of central processing, i.e. processing all antenna elements together, and propose local approaches.   
%This leads shifting of directions of arrival of signals along the array. Second, due to different directions and obstacles between the users and the array visible part of VLA is different for each user. Thus, some parameters such as average received power varies along the array axis. 
%It is worth mentioning that in \cite{measurement_paper}, non-stationary behaviour of received power were observed through relatively compact arrays.
Note that the vast majority of the existing works on massive MIMO are based on conventional standard models with stationary characteristics of the channel \cite{marzetta}. In \cite{emil}, an information theory study on non-wide sense stationary 
characteristics of massive MIMO channels is available where different parts of the array see different propagation paths.
The problem of user assignment in large intelligent surfaces is studied in \cite{Husha} in an interference-free environment. 
%Moreover, they assume a less crowded scenario which keeps the number of antenna arrays to be more than the number of users. However, we deal with more general model where the interference has an important impact on the system performance.
%Therefore, we utilize an interference cancellation method to get even better efficiency.
%Furthermore, we cope with crowded scenarios where the number of users is more than the antenna arrays and the model in \cite{Husha} fails to assign users.

To exploit cluster visibility regions, we propose new algorithms for uplink data detection. 
One of the challenges in xMaMIMO is its practical implementation, especially the enormous computational load that is required. 
To reduce the computation load of the system, we divide the array into smaller, disjoint units, referred to as subarrays and we distribute the computations among them. 
Then, we propose two types of uplink receivers.
The first receiver is based on distributed linear data fusion (DLDF), where the users are first softly detected per subarray and then linearly fused in a centralized manner to produce the final soft information used to reconstruct the symbols.
%Hence, we consider applying successive interference cancellation (SIC) on top of DLDF, which leads to substantial performance improvement, while still maintaining low computational complexity.
Next, relying on the non-stationary nature of the xMaMIMO system and drawing inspiration from coded random access, we propose a decentralized receiver of very low complexity where processing is executed locally per subarray with the fusion centre acting only as a forwarding node, relaying messages among the subarrays. One important factor here is the order of local processes and our proposed method copes delicately with it.
The simulation results confirm the practical potential of the proposed receivers for xMaMIMO systems especially in crowded applications.

\section{System Model}

%\subsection{Preliminaries}
We consider an xMaMIMO system. 
As discussed earlier, such infrastructure can be deployed along walls of buildings in urban sprawls, airports, shopping malls, even stadiums and they are envisioned to provide services to massive crowds.

A possible way to deal with the enormous computational load of the xMaMIMO system is to distribute the computation within separate processing units, referred to as \textit{subarrays}.
Depending on the specific implementation of the system and the actual physical constraints, a subarray can be defined in various ways.
For instance, a subarray can correspond to a separate physical component.
To see this, consider a large stadium.
To provide high quality connectivity, an xMaMIMO system can be deployed along its walls.
Depending on the actual deployment burden and cost, the operator might choose to mount individual arrays and connect them into a central processing unit using a cloud radio access network architecture. In such case, the number and the sizes of the subarrays is fixed.
Alternatively, the operator might install a single array and provide logical interconnections between different portions of it. 
Here, the subarrays can be defined flexibly, adapting their size, number and position to the evolving data traffic conditions.
We note that our framework is applicable to both cases as well as any combination in-between.
%The concept of sub-arrays is not just a theoretical fact since we can use it in practical implementation of the VLAs. For instance, consider a large stadium with a capacity of about 50K people. In order to provide high quality cellular connectivity we can implement a linear antenna array around the stadium's walls. Regarding difficulties in hardware implementation of whole array at once, we should carry out smaller components (as sub-arrays) and try to interconnect them in a central processing unit.  

%(THIS SHOULD COME LATER IN THE DETECTION ALGORITHMS SECTIONS SINCE IT IS NOT RELEVANT FOR THE SYSTEM MODEL)We assume a perfect Channel State Information (CSI) for each sub-array.
%In order to model the problem of user data detection we try to define bipartite graph of the system. Then, we propose an algorithm to detect users based on sub-array processing. 
%\subsection{Baseband signal model}

Let $M$ and $K$ denote the number of antennas and simultaneously active users, respectively.
We assume narrow-band transmissions; 
$\mathbf{x}\in\mathbb{C}^K$ denotes the vector of complex input symbols, $\mathbf{H}\in\mathbb{C}^{M\times K}$ is the complex channel matrix and $\mathbf{n}\sim \mathcal{CN}(0,\sigma_n^2\mathbf{I}_M)$ is the AWGN ($\mathbf{I}_M$ denotes the identity matrix). We model the received baseband signal $\mathbf{y}\in\mathbb{C}^{M}$ across the whole array as follows:
\begin{align}\label{eq:general_model}
\mathbf{y}=\mathbf{H}\mathbf{x}+\mathbf{n}.
\end{align}
Let $\mathbf{h}_k$ denote the $k$-th column of $\mathbf{H}$, corresponding to user $k$;
%Also, channel matrix can be written as
%$ \mathbf{H}=[\mathbf{h}_1\:\mathbf{h}_2\dots\mathbf{h}_K] $ with $ \mathbf{h}_k $ denoting the channel for user $ k $.
in this work, we adopt the following channel  model \cite{larsson_energy}:
\begin{align}\label{eq:ch_model}
   \mathbf{h}_k=\sqrt{\mathbf{w}_k} \odot\bar{\mathbf{h}}_k,
\end{align}
with $\odot$ denoting the element wise (Hadamard) products between two equal-size vectors.
$\mathbf{w}_k$ captures the effect of large scale fading which in turn is a function of the distance of the user from the array, denoted with $\mathbf{d}_k$, and the propagation properties of the environment; here, we employ the following simplified propagation model \cite{tse}:
\begin{align}
\mathbf{w}_k=\beta \mathbf{d}^{\gamma}_k,
\end{align}
where $\beta$ is a attenuation coefficient \cite{tse} and $\gamma$ is the pathloss exponent.
 $\bar{\mathbf{h}}_k \sim \mathcal{CN}(0,\mathbf{I}) $ accounts for fast fading.
%This large scale models the geometric attenuation and shadow fading and changes very slowly with time.
%Here, we use \begin{align}
%\mathbf{w}_k=\beta \mathbf{d}^{\gamma}_k
%\end{align}
%for large scale fading with $\mathbf{d}_k$ is the distance vector between user $k$ and array elements and $\beta$ is a coefficient depending on the environment \cite{tse} and $\gamma$ is pathloss exponent. Moreover, small scale fading follows $\bar{\mathbf{h}}_k \sim \mathcal{CN}(0,\mathbf{I}) $.

We split the xMaMIMO system into $B$ subarrays, each with $M^{(b)}\geq K,b=1,\hdots,B$ antennas such that $\sum_{b=1}^B M^{(b)}=M$; the received signal per subarrays is denoted by $\mathbf{y}^{(b)}\in\mathbb{C}^{M_b}$ and can be written as:
%For a distributed system with $ B $ subarrays and each with $ M_b $ antennas, we have below equation for each $ b=1,2,\dots,B $, unit as,
\begin{align}\label{eq:subarray_model}
\mathbf{y}^{(b)}=\mathbf{H}^{(b)}\mathbf{x}+\mathbf{n}^{(b)},
\end{align}
for any $b=1,\hdots,B$.
Without loss of generality, in the rest of the paper we will assume that all active users transmit with equal power ($E|x|^2 = 1$).
%Knowing $ \mathbf{H}=[\mathbf{H}^{(1)T}\:\mathbf{H}^{(2)T}\dots\mathbf{H}^{(B)T}] $.

%\subsection{Detection Methods}

%\subsubsection{Successive Interference Cancellation}
%Finally, we will also consider Successive Interference Cancellation (SIC).
%This nonlinear method is based on iterative decoding of the users. The principle of SIC is to decode one user data at each iteration and remove its contribution from the received signal. A simple description of SIC is as the following:
%\begin{itemize}
%    \item At each iteration, one user data is decoded considering the other users as interference.
%    \item The decoded data is then removed from the received signal.
%    \item In the next iteration, the selected user is decoded with one less interferer, providing accurate and simpler decoding.
%\end{itemize}
%The most important difference between the SIC method in SA and what we use here is that in SA it is impossible to recover collision slot when there is no prior information about the contributing users. However, here and in MIMO array even without singletons, there is still a chance to completely decode the received signal.

\section{Multiuser Detection Algorithms}
\label{algorithms_sec}
In this section, we develop algorithms for multiuser symbol detection in xMaMIMO systems.
Throughout, we assume perfect Channel State Information (CSI) at the receiver.

\begin{figure}
	\centering
 	\includegraphics[width=1\linewidth,trim={0cm 1.3cm 0cm 1.2cm },clip]{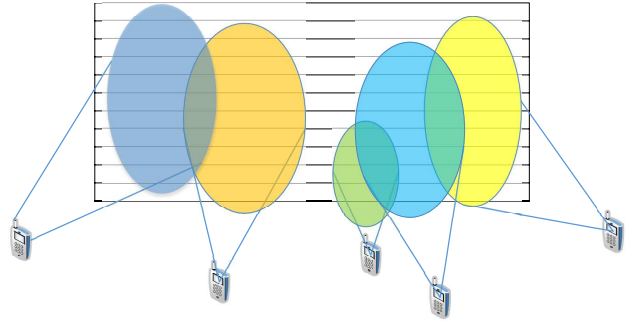}
	\caption{ \small An example of extremely large M-MIMO array with spatial non-stationary regions along the array. Each user has a specific visibility region according to the channel conditions.
	}
	\label{fig:ex1}
\end{figure}

We distinguish between two different regimes of operation of the system: (i) stationary regime, where we assume that the users' energy spread across the whole array (in other words, each user ``sees'' the whole array), and (ii) non-stationary regime, where we assume that the energy of each user is predominately concentrated on a limited number of antennas (see Fig. \ref{fig:ex1}), which is usually significantly smaller than $M$ (i.e., each user ``sees'' only limited portion of the array).
Obviously, the inherent, natural regime of operation of the system would be the non-stationary one in general, since the uplink power of each user will be unevenly distributed along the antenna array.
Here, the distinction between the two regimes is done according to the knowledge of the receiver, i.e., when we say stationary regime, we mainly refer to the aspect of receiver agnosticism towards the non-stationary nature of the system.

\subsection{Stationary regime}

\subsubsection{Centralized Zero-Forcing Receiver}

Given the main underlying assumption, i.e., each user ``reaches'' every antenna of the array, a straightforward way to perform multiuser symbol detection is to process the complete received signal $\mathbf{y}$.
This can be done simply via Zero-Forcing (ZF); specifically, the ZF receiver for user $k$, denoted by $ \mathbf{F}_{ZF,k} $, can be written as follows:
\begin{align}
\mathbf{F}_{ZF,k}[\mathbf{H}]=\frac{\mathbf{h}^H_k\mathbf{P}^{\bot}_{\mathbf{\bar{H}_k}}}{\mathbf{h}^H_k\mathbf{P}^{\bot}_{\mathbf{\bar{H}_k}}\mathbf{h}_k},
\end{align} 
with $\mathbf{P}^{\bot}_{\mathbf{\bar{H}_k}}=\mathbf{I}-\mathbf{\bar{H}_k}(\mathbf{\bar{H}_k}^H\mathbf{\bar{H}_k})^{-1}\mathbf{\bar{H}_k}^H$ \cite{mimo_book}; $\mathbf{\bar{H}_k} $ is obtained from $ \mathbf{H} $ by removing its $k^{th}$ column $\mathbf{h}_k$.
%{For simplicity and for the rest of the paper, we denote generically as $\mathbf{F}_{ZF,k}$ the ZF receiver adapted to the channel associated to the received signal considered.}
The post-processing SNR of the ZF receiver obtains the following form:
\begin{align}
\text{SNR}_{ZF,k}=\rho \, \mathbf{h}^H_k\mathbf{P}^{\bot}_{\mathbf{\bar{H}_k}}\mathbf{h}_k,
\label{snrzf}
\end{align}
with $ \rho={1}/{\sigma_n^2}$.
Given the extreme dimension of the aperture and potentially the extremely crowded setup, one should immediately note the computational burden of the centralized scheme.
To reduce the computational complexity, we propose two schemes based on subarray processing.
In both cases, the underlying idea is simple; instead of processing $\mathbf{y}$ fully, first process $\mathbf{y}^{(b)},b=1,\hdots,B$ and then perform linear soft fusion in a centralized manner.

\subsubsection{Distributed Linear Data Fusion Receiver}
\label{linearfusion}

We introduce a simple, distributed linear data fusion (DLDF) method that combines softly the detected signals from each individual subarray. Furthermore, soft information of each user is obtained by
\begin{align}
    \hat{x}^{(b)}_k=\mathbf{F}_{ZF,k}[\mathbf{H}^{(b)}]\mathbf{y}^{(b)}
    \label{x_soft}
\end{align}
For each user $ k $ we define the combined DLDF symbol $\hat{x}_k$ as follows:
\begin{align}
\hat{x}_k&=\sum_{b=1}^{B}\alpha^{(b)}_k\hat{x}^{(b)}_k,\label{datafusion}%\\
%\text{MSE}_k&=\sum_{b=1}^{B}\alpha^{(b)2}_k\text{MSE}^{(b)}_k,
\end{align}
where $ \alpha^{(b)}_k $ is the weight for user $ k $ using from subarray $b$; note that $ \sum_{b=1}^{B}\alpha^{(b)}_k=1 $. 
%(DEFINE THE MSE IN A SEPARATE EQUATION).
It is worth mentioning that \eqref{datafusion} is done in the central unit after receiving all the soft information from the subarrays.
Also, mean squared error (MSE) of each user on subarrays is defined as:
\begin{align}
    \text{MSE}^{(b)}_k= E |{x}^{(b)}_k-\hat{x}^{(b)}_k|^2
\end{align}
where $E$ denotes the expectation operation. Here, it is taken with respect to the noise. 

As the noise is assumed independent across subarrays, the overall MSE when data fusion is performed is:
\begin{align}
\text{MSE}_k&=\sum_{b=1}^{B}\alpha^{(b)2}_k\text{MSE}^{(b)}_k,
\end{align}
The objective is to minimize  
$\text{MSE}_k$ with the constraint $\sum_{b=1}^{B}\alpha^{(b)}_k=1$.
%we therefore, formulate the following constrained optimization problem:
%\begin{align}
 %   &\text{arg}\min_{ \mathbf{\alpha}_k}\sum_{b=1}^{B}\alpha^{(b)2}_k\text{MSE}^{(b)}_k\\
    %&\text{Subject to:} \sum_{b=1}^{B}\alpha^{(b)}_k=1\nonumber
%\end{align}
Using the Lagrange multiplier method \cite{boyd} gives us the optimal weights:
\begin{align}
\alpha^{(b)2}_k=\frac{\frac{1}{\text{MSE}^{(b)}_k}}{\sum_{b=1}^{B}\frac{1}{\text{MSE}^{(b)}_k}},
\end{align}
for $b=1,\hdots,B$.
Given that all users use equal transmit power, normalized to $1$, we have that $ \text{SNR}^{(b)}_k= {1}/{\text{MSE}^{(b)}_k}$, which yields:
\begin{align}
\alpha^{(b)}_k=\frac{\text{SNR}^{(b)}_k}{\sum_{b=1}^{B}\text{SNR}^{(b)}_k}
\label{alpha2}
\end{align}
for $b=1,\hdots,B$.
The complete algorithm is summarized in Algorithm \ref{alg1}.

%\subsubsection{SIC-empowered DLDF}

%After DLDF has softly combined the symbols from the subarrays, the receiver can further boost the performance by applying conventional successive interference cancellation (SIC) to the fused signal.

% The scheme can be seen as an enhancement of the DLDF via SIC-based empowerment and is only practical if the computational budget allows the computational burden of the SIC.
%Note that, due to the per subarray processing manner of DLDF and SIC-empowered DLDF, complexity is much less than the centralized-ZF where all of the computations are done based on the whole set of array elements.
%This method is consist of two parts: a local part where subarrays extract soft information about each user, then, the second part where central unit performs data fusion mentioned in \eqref{datafusion} and applies conventional SIC.

\begin{algorithm}[t]
%\hspace{0.5cm}
	\SetAlgoLined
	\KwResult{Estimates of $x_k,k=1,\hdots,K$}
	%\KwResult{Average BER of the detection system }
	\emph{Initialize:} $ \mathbf{H},\:K,\:B,\:M^{(b)},\:\mathcal{K}=\left\{1,\hdots,K\right\}$
	
	\textbf{Stage I:} \emph{Distributed Linear Data Fusion (DLDF)}\\
	1. compute $\hat{x}^{(b)}_k,k\in\mathcal{K},b=1,\hdots,B$ via \eqref{x_soft}\\
	2. compute $\alpha_k^{(b)},k\in\mathcal{K},b=1,\hdots,B$ via \eqref{alpha2}\\
	3. compute $\hat{x}_k,k\in\mathcal{K}$ via 
	\eqref{datafusion}\\
	4. perform hard decision over $\hat{x}_k,k\in\mathcal{K}$ and terminate
	% with $\alpha^{(b)}_k $ mentioned in \eqref{alpha2} \;
	%Sort the users according to $\sum_{b=1}^{B}\text{SNR}_{ZF}^{(b)}$ in \eqref{snrzf} for each user and select the one maximizing it (${k}^\ast$)\;
	%Hard decision over $ \sum_{b=1}^B\hat{x}^{(b)}_{k^{\ast}}$ which gives $ \tilde{x}_k^{\ast}$\;
	%Reconstructing the transmitted signal $ \mathbf{h_k}\tilde{x}_k^{\ast}$ and remove it's effect from received signal\;
	%$ D=D\setminus k^{\ast} $ (one user is detected)\;
	%$ n=K-1 $\;

	\caption{\small  Distributed Linear Data Fusion receiver. }
	\label{alg1}
\end{algorithm}

\subsection{Non-stationary regime}

In this case, we assume that the users have a limited visibility region of the array, which is illustrated in Fig.~\ref{fig:ex1}.
Hence, the non-stationary regime of operation can be seen as a special case of the stationary one, implying that we can easily apply any of the receivers described in the previous subsection.

Nevertheless, we introduce a simple method, inspired from the concept of coded random access in slotted aloha IoT systems.
They key idea operates as follows: given the non-stationary nature of the array, each user is predominantly present only on very limited number of subarrays, i.e., all of its power is concentrated over limited number of subarrays.
As a result, the system becomes inherently \emph{sparse}, implying that the subarrays where a user is not present should not be processed for that specific user.
So, in principle, we can obtain a sparse bipartite graph, representing the connections of the users to the subarrays after which we can apply the principles of successive elimination of connections from the graphs as in coded random access.
This further reduces the computational cost but since we are neglecting some portion of the signal energy, $p_0$, and treat it as interference at the remaining arrays, and we do not perform any soft fusion at the central processing unit, it is reasonable to expect that the performance of the method might be slightly degraded in some configuration regions of the system (i.e., specific values of $K$ relative to $B$ and $M_b,b=1,\hdots,B$).

\begin{algorithm}[t]
	\SetAlgoLined
	\KwResult{$\mathcal{H}$}
	\emph{Initialize:} $ \mathbf{H},\:K,\:B,\:M_b,\: p_0,\: \mathcal{H}=\{0\}^{B\times K},\:\mathcal{B}=\left\{1,\hdots,B\right\}$ \\
\For{$k = 1$ to $K$ }{
1. reinitialize $p_k=0$

2. compute total cumulative power $P_k$

3. compute per subarray power $P_k^{(b)},b\in\mathcal{B}$

	\While{$ p_k\leq p_0\cdot P_k $}{
		1. find $b^*=\max_{b\in\mathcal{B}} P_k^{(b)}$
		
	    2.	$p_k=p_k+P_k^{(b^*)}$
	
	    3.	set $\mathcal{H}(b^*,k)=1$
	
	    4.	$\mathcal{B}=\mathcal{B}\setminus {{b}^{*}}$ %$P_k(max_{\text{index}})=\phi$
	
		}
		}

	\caption{\small Bipartite graph construction from $\mathbf{H}$.}
	\label{alg3}
\end{algorithm}

The most attractive feature of the proposed method is the fact that the bipartite graph can be constructed very simply, exploiting the sheer fact that the receiver has perfect CSI, i.e., it knows $\mathbf{H}$; 
in other words, by observing the $k$-th column, the receiver can determine which parts of the array the dominant part of the power of user $k$ is allocated.
%A simple procedure to do that is the following: (i) the receiver fixes a threshold $p_0<1$ on the cumulative received power per user; for instance, $p_0$ can be $0.9$, implying that we want to identify the sub-arrays where $90\%$ of the total power of each user is concentrated, (ii) the receiver computes the cumulative power of each user per sub-array, (iii) the receiver orders the sub-arrays per user in decreasing order w.r.t. the cumulative power, (iv) the receiver sums the accumulates the cumulative power per user across sub-arrays, and in each step compares the total cumulative power with $p_0$; the summation stops once the threshold is surpassed.
This way, the receiver obtains a binary matrix $ \mathcal{H} \in \{0,1\}^{B\times K}$. We use $ \mathcal{H} $ to construct a bipartite graph. 
Note that, at the beginning of the algorithm 3, the central unit runs Algorithm 2 according to the CSI and then sends the order of the detection to each subarray. This means that each subarray receives a schedule consisting of the list of users that should be detected by that subarray. This is the only centralized broadcasting in this algorithm and the rest of it is decentralized. 

\begin{figure}
	\centering

	\includegraphics[width=1\linewidth]{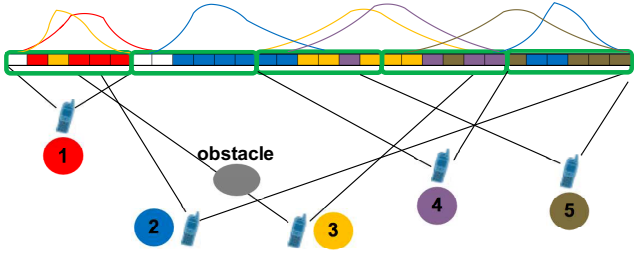}
	\caption{ \small An example of linear M-MIMO array with different user visibility regions. Equivalent graph representation for this system is shown in Fig. \ref{fig:ex2} (a).}
	\label{fig:ex3}
	
\end{figure}

\begin{figure}
	\centering

	\includegraphics[width=1\linewidth]{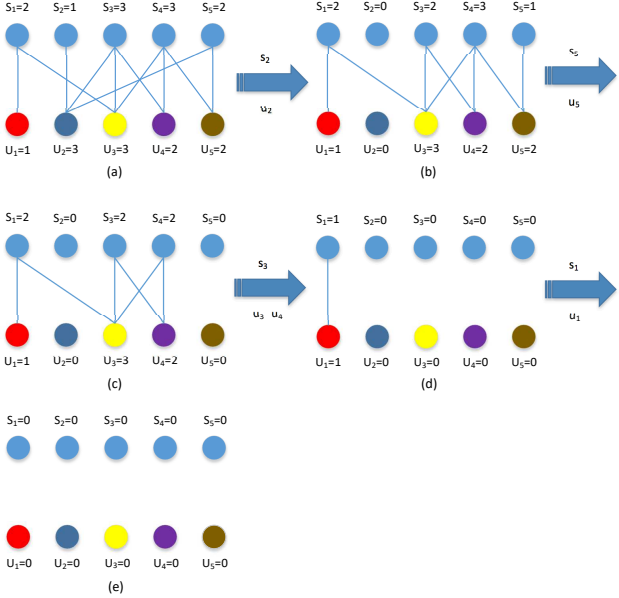}
	\caption{ \small An example of the proposed detection model on a bipartite graph model with $K=5$ users and $B=5$ subarrays. (a): connections of the users to each array. The subarray with the lowest number of users is selected ($S_2$). The corresponding user symbols are detected and removed from the other subarrays. (b): the procedure is repeated for $ S_5 $ and $ U_5 $. (c): $ S_3 $ is randomly selected. ZF detection is used to decode user 3 and user 4. Their data is removed from the other subarrays. (d): the last user is detected. 
	}
	\label{fig:ex2}
	
\end{figure}

The procedure  is described in Algorithm~\ref{alg3}. Moreover, an example of xMaMIMO with 5 users is illustrated in Fig.~\ref{fig:ex3} where the energy distribution of each user on antenna arrays is also presented. The equivalent bipartite graph representation of this setup is shown in Fig.~\ref{fig:ex2} (a).
The extension of this binary graph to a weighted one, where the weights show the portion of user's total power, is left for future work.


The bipartite graph constructed this way is characterized by the following quantities: (i) a set $ \mathcal{B} $ of $ B $ nodes representing subarray units, (ii) a set $ \mathcal{K} $ of $ K $ user nodes and, (iii) a set $ \mathcal{E} $ of edges, i.e. connection between users and subarrays.
We use $\mathcal{G} = (\mathcal{B} , \mathcal{K} , \mathcal{E}) $ to denote this graph.
We also define the node degrees $ S_b,b\in\mathcal{B} $ and $ U_k,k\in\mathcal{K}  $; the degrees give the number of edges connected to each of the nodes in $\mathcal{B}$ and $\mathcal{K}$, respectively.
For instance in Fig. \ref{fig:ex2}(a), subarray $b=1$ only receives signal from user $k=1$ and user $k=3$; therefore, its degree is $ S_1 = 2 $.

Once the graph has been constructed, we apply simple symbol detection strategy inspired from coded random access \cite{sparse}.
Hence, we search for subarrays with the lowest number of users; we detect the symbols of those users and subsequently remove them from the other subarrays.
%The support of the problem in coded random access is the set of integers, and the algorithms for successive removal of users to resolve collisions fails to perform is no singletons are detected.
%Oppositely, our problem is inherently continuous, and even if no singletons, i.e., subarrays where only one user is present, we can still recover the signal. 
An illustrative example for the procedure is shown in Fig.~\ref{fig:ex2}.
We assume xMaMIMO system with $K=5$ users and $B=5$ subarrays with graph representation shown in Fig.~\ref{fig:ex2}(a).
After computing the degrees, we see that $S_2 = 1$.
Thus, we start signal detection in subarray $ b=2 $ for user $ k=2 $.
Then, we remove any other edges corresponding to user $k=2$ from the graph.
We repeat the procedure for $ S_5 $ and $ U_5 $ in Fig.~\ref{fig:ex2}(b). In step (c), we have three nodes with same degree and similar conditions. We randomly choose $ S_3 $ and start a ZF detection within subarray $b=3$ between users $k=3$ and $k=4$.
After recovering both of them,  we remove all edges from the graph corresponding to those users.
Finally, in part (d) we have a singleton node that can be easily detected. Now, since all the users are detected, the algorithm terminates.
The complete algorithm is summarized in Algorithm \ref{alg2}.

\begin{algorithm}[t]
	\SetAlgoLined
	\KwResult{Estimates of $x_k,k=1,\hdots,K$}
	\emph{Initialize:} $ \mathbf{H},\:K,\:B,\:M_b$, 
	 $ \mathcal{B}=\{1,\hdots, B\},\: \mathcal{K}=\{1,\hdots, K\}  $
	
	1. compute $ \mathcal{H}$ via \textbf{Algorithm}~\ref{alg3}
	
	\While{$ \mathcal{K}\neq\emptyset $}{
		1. compute node degrees $S_b,b\in\mathcal{B}$, $U_k,k\in\mathcal{K}$
		
		2. find $b^* = \min_{b\in\mathcal{B}}\left\{S_b\right\}$

	\uIf{$b^*= 1 $(only user $k^{*}$in the subarray with minimal degree)}
	{
		1. compute $\hat{x}_{k^{*}}^{(b^*)}$ via \eqref{x_soft}  
		
		%2. compute $\tilde{x}_{k^{*}}^{(b^*)}$ via hard decision over $\hat{x}_{k^{*}}^{(b^*)}$
		
		2. broadcast $ \hat{x}_{{k}^{\dagger}}$ so other subarrays remove it from $\mathbf{y}^{(b)}$ for $b\in \mathcal{B}\setminus b^* $
		
		3. $\mathcal{K} = \mathcal{K}\setminus {k^{*}}$
	}	
	
	\uIf{$b^* >1 $(multiple users $\mathcal{K}^*\subset\mathcal{K}$ in the subarray with minimal degree)}
	{
		\While{$\mathcal{K}^*\neq\emptyset$}{
	
	    1. sort the users according to $\text{SNR}_{ZF,k},k\in\mathcal{K}^*$ in \eqref{snrzf}
		
		2. find $k^{\dagger}=\max_{k\in\mathcal{K}} \text{SNR}_{ZF,k}$
		
		%3. compute $ \tilde{x}_{{k}^{\dagger}}$ via hard decision over $ \hat{x}_{{k}^{\dagger}}$
		
		3. broadcast $ \hat{x}_{{k}^{\dagger}}$ so all subarrays remove it from $\mathbf{y}^{(b)}$ for $b\in \mathcal{B}$
		
		4. $ \mathcal{K}^*=\mathcal{K}^*\setminus {{k}^{\dagger}} $

	}
	
	6. $\mathcal{K}=\mathcal{K}\setminus {\mathcal{K}^*}$

	}
		
		}

	\caption{\small Low complexity multiuser detection in non-stationary regime}
	\label{alg2}
\end{algorithm}

\section{Complexity, Convergence and Delay Analyses}
In this section we first consider computation complexity comparison between the proposed algorithm and the linear data fusion using ZF detector. Convergence and delay characteristics of the proposed algorithms are discussed next.

\subsection{Complexity of DLDF}
In DLDF we have three phases for user detection which have the following complexities:
\begin{enumerate}
    \item Data detection: Consists of ZF matrix inversion for all users in all of the subarrays with $M_b K$ elements, which have a complexity order of
    $B K( K)^3$.
    \item SNR extraction: This function is also for all users in all of the subarrays containing 
    $B K(M_b (K-1))$.
    \item Soft fusion: The last part with $B K$ matrix multiplications.
\end{enumerate}
% \subsection{Complexity of Algorithm 1}
% The analyses for this part is similar to the DLDF case with one additional step for the SIC scheme, which runs over $K$ users over equivalent array of $M_b$ antennas. Therefore, it consists of the following number of operations:

\subsection{Complexity of Algorithm 3}
We study this part with two extreme cases that could happen regarding the nature of non-stationarity.
\subsubsection{Worst case}
This case occurs when we have all of the users in all of the subarrays or when we have them in only one subarray. Thus, the algorithm sets $\min N_s=K$ and performs detection over only one subarray.
Moreover SNR extraction is also done for all users in this subarray containing 
    $K(M_b (K-1))$. Therefore,
the complexity of this part is at most $ K^4+K(M_b (K-1))$ calculations. 
\subsubsection{Best case}
This case happens when we have users evenly distributed between subarrays meaning that each subarray performs detection over $\frac{K}{B}$ users. Therefore the complexity of the detection part is
\begin{align}
    B  \left[\left(\frac{K}{B}\right)^3+\left(\frac{K}{B}-1\right)^3+\dots+1\right]<\frac{K^4}{B^3}.
\end{align}
Also, for the ordering part we have $\frac{K M_b K}{B}$ computations.
TABLE \ref{table} provides the complexity comparison between the aforementioned algorithms. 

A rough comparison between the complexities of the Algorithms reveals that the complexity reduction of Algorithm 3 scales with $B$ in the worst case and with $B^4$ in the best case.
If $B$ is of the order of $10$, we see that significant computational power can be saved by employing Algorithm 3.

\begin{table}
\centering
\caption{Complexity comparison of the studied methods}
\begin{tabular}{c c} 
 \hline
 Methods & Number of multiplications  \\ [0.5ex] 
 \hline
 \vspace{0.3cm}
 ZF-DLDF & $B K( K)^3+B K( (K-1))+B K$  \\ 
 [2ex] 
 
 {Algorithm 3} & Worst case:  $ K^4+K(M_b (K-1))$ \\
 [1ex]
 &Best case:  ${K^4}B^{-3}+\frac{K M_b K}{B}$\\
 
 \hline
 \end{tabular}
 \label{table}
\end{table}

\subsection{Convergence Analyses}
The analyses for the convergence for Algorithm 3  over graphical model can be found in \cite{sparse}. Moreover, considering this algorithm as an extension to the model in \cite{sparse} by enabling continuous signal space, i.e. we recover a part of data at each step even on non-singleton nodes, it will converge even faster due to this claim.

\subsection{Delay Characteristics}
In order to highlight the trade-off between using our proposed algorithms instead of the centralized methods, we discuss delay properties of the algorithms.  Algorithm~\ref{alg2} introduces some delay due to the sequential nature of the algorithm. 
For example, while a selected subarray $b$ performs the detection of a given symbol, the other subarrays carrying the same symbol should wait for the input from subarray $b$ to perform interference cancellation. 

This waiting time vanishes when we have sparse channels since different subarrays become independent as they involve different users. Thus, they can work together in a parallel mode.  A deeper analysis of the overall delay is left for future work. 

\section{Simulation Results}

In this section we compare the Bit Error Rate (BER) performance of the proposed algorithm and DLDF. We assume a linear xMaMIMO configuration (see Fig. \ref{fig:ex3}) for our simulations setup. We use Monte-Carlo simulations to generate the channel realizations. Some of the fixed variables are: $\beta=1$, $\gamma=2$, array length $=100$ m, power threshold for constructing the bipartite graph $p_0=0.9$ and we use an 8-PSK input constellation.  
Moreover, we assume that the user location has a random uniform distribution along the array. 
Note that user distribution and antenna array length have a direct impact on the large scale fading characteristics and therefore control the bipartite graph structure.
% Fig. \ref{fig:sim3} compares the BER performance between ZF-DLDF, centralized ZF and Algorithm 1 versus the pre-processing SNR $\rho$ for different $B$ and $M_b$ values. In centralized ZF we assume that all of the subarrays are working together and the central processing unit is in charge of all computations. As it can be seen,  Algorithm 1 performs better than DLDF in all SNR values. The cost we pay here is the complexity of subarray SIC which improves the detection accuracy. Moreover, with reducing the number of subarrays while the total number of antenna $M$ is fixed, we see that the performance of Algorithm~\ref{alg1} surpasses the centralized scheme.

%In Fig.~\ref{fig:sim1}, we compare the performance of Algorithm 3 and DLDF with respect to the number of users in the system and also the number of subarrays $B$. As it can be seen, the error rate of Algorithm~\ref{alg2} is always better than DLDF for any number of users. Furthermore, increasing the number of subarrays, while $K/M_b=1$ is fixed, does not have significant effect on DLDF while in Algorithm~\ref{alg2} it improves the performance. The reason for this is that with larger $B$ we have higher resolution for graph representation. Thus, the interference cancellation method implemented in the algorithm deals better with the interfering effect of the increased number of the users. This also means that the ordering of the subarrays to perform detection perfectly adapts itself with much higher graph dimensions.

Fundamentally, for a given average user load per subarray, the BER performance is determined by the dimension of the visibility region seen by each user. Here, the users are at the same distance and uniformly distributed along the array, so that the average size of the visibility region is the same per user (except for users at the edges which have a non-significant impact for a large enough number of users).
In the simulation, we study the following factors: 
%a) the dimension of the array, b) the number of antenna per subarray, c) the number of subarrays, d) the total number of users.
a)  the number of subarrays, b) the total number of users and c) the number of antenna per subarray.

First, we compare the BER of the detection methods for different numbers of subarrays, $B$, while the number of antennas is fixed, $M=512$, starting from a number of 2 subarrays (256 antennas per subarray) and ending by a number of 32 subarrays (16 antennas per subarray). We observe: 
\begin{itemize}
    \item Algorithm 3 performs significantly better than the DLDF.
   \item 	Linear processing: when the number of antennas per subarray is asymptotically large, due to the law of large numbers, the processing decouples across the subarrays (this fact is supported by the  rich scattering assumption).
   Subarray processing becomes equivalent to a centralized linear processing. We observe an almost stable performance level (corresponding to the centralized processing) until a number of antennas per subarray smaller then 32. 
   \item 	Algorithm 3: the performance of algorithm 3 is degraded when the number of subarray is small. The reason is that the algorithm lacks degrees of freedom for the SIC mechanism to have its full effect. Performance saturates when the resolution offered by the number of subarrays reflects the non-stationarity patterns, more precisely when each subarray offers a stationary picture of the received signal.
\end{itemize}

In Fig.~\ref{fig:sim1}, we compare the performance of Algorithm 3 and DLDF with respect to the number of users. The total array size is kept fixed. The array is comprised of $B$ subarrays, each with a number of antennas $M_b=K$. Note that the subarrays are not adjacent in this simulation. We make the following observations:
% to the array dimension, the number of users in the system and number of subarrays $B$.
\begin{itemize}
   \item Again,  algorithm 3 performs significantly better than the DLDF.
 \item As the number of antennas grows at the same speed as the number of users,  the user load per subarray is maintained so that performance remains approximately constant.
 
 \item We observe an improvement of algorithm 3 as the granularity increases, i.e. the number of subarrays.

 %   \item Array dimension/number of users: we maintain the ratio $K/M_b=1$ fixed. Therefore, the array dimension grows at the same speed as the number of users. The user load per subarray is maintained and, as the average size of the visibility regions remains the same, performance remains approximately constant.
%\item	Number of subarrays: as mentioned above, we observe that the performance of DLDF remains constant.  This indicates that the processing per subarray is decoupled, as the number of antennas per subarray is large. 
%Furthermore, we observe an improvement of algorithm 3 as the granularity increases. 
\end{itemize}

\begin{figure}
	\centering
	\includegraphics[width=1\linewidth,trim={2cm 0 1.6cm 0 },clip]{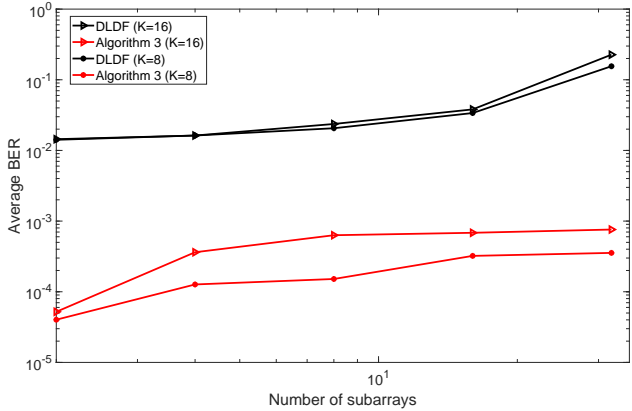}
	\caption{ \small The effect of the number of subarrays $B$ on the average BER of the detection systems. We set $M=512$ and SNR$=25$dB.}
	\label{fig:sim2}
	
\end{figure}

\begin{figure}
	\centering
	\includegraphics[width=1\linewidth,trim={2cm 0 2cm 0 },clip]{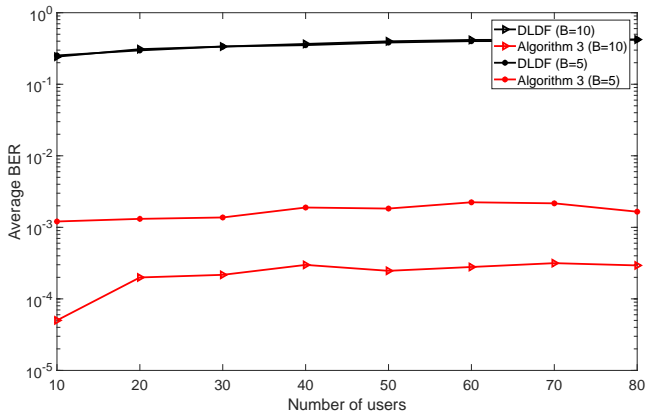}
	\caption{ \small Performance comparison between Algorithm 3 and DLDF with respect to number of active users. SNR$=25$dB and $K/M_b=1$.}
	\label{fig:sim1}
	
\end{figure}

\section{Conclusions}

In a massive MIMO systems with extremely large arrays, users can effectively communicate only  with a sub-part of the array called a visibility region. A receiver design should be adapted to this kind of non-stationary patterns with partially overlapping visibility regions. The receiver architectures proposed in this paper are based on subarray processing where part of the computational load is carried out. A central unit coordinates the operations at each subarray and proceeds to data fusion.  
%In this paper we studied multiuser data detection in massive MIMO systems with extremely large arrays. We considered two scenarios with and without wide sense stationary conditions on the channel.
We proposed a linear data fusion method, as well as a graph-based algorithm inspired from coded random access which uses low complexity and distributed scheme for the data detection. This method converts the propagation environment of the channel into a bipartite graph and detects the users in a novel scheme.

\bibliographystyle{IEEEtran}
%%\bibliography{bib}

\end{document}